**Accelerated calculation of configurational free energy using a combination of reverse Monte Carlo and neural network models: Adsorption isotherm for 2D square and triangular lattices**

Akash Kumar Ball[a], Swati Rana[b], Gargi Agrahari[a] and Abhijit Chatterjee[a]*

[a] Department of Chemical Engineering, Indian Institute of Technology Bombay, Mumbai 400076, India

[b] Department of Energy Science and Engineering, Indian Institute of Technology Bombay, Mumbai 400076, India

*Email: abhijit@che.iitb.ac.in

**ABSTRACT**

We demonstrate the application of artificial neural network (ANN) models to reverse Monte Carlo based thermodynamic calculations. Adsorption isotherms are generated for 2D square and triangular lattices. These lattices are considered because of their importance to catalytic applications. In general, configurational free energy terms that arise from adsorbate arrangements are challenging to handle and are typically evaluated using computationally expensive Monte Carlo simulations. We show that a combination of reverse Monte Carlo (RMC) and ANN model can provide an accurate estimate of the configurational free energy. The ANN model is trained/constructed using data generated with the help of RMC simulations. Adsorption isotherms are accurately obtained for a range of adsorbate-adsorbate interactions, coverages and temperatures within few seconds on a desktop computer using this method. The results are validated by comparing to MC calculations. Additionally, H adsorption on Ni(100) surface is studied using the ANN/RMC approach.

**1. Introduction**

In this current era of artificial intelligence, machine learning (ML) based models are increasingly being used in molecular simulations. ML has been applied to a variety of problems ranging from quantum mechanical calculations[1], to the construction of classical interatomic potentials[2–5], structure-prediction and materials design[6,7], and predicting thermo-physical properties[8] and activation barriers[9]. An artificial neural network (ANN) model is often used in these applications. The popularity of ANN arises from its general functional form, which enables users to train the ANN model to high-dimensional nonlinear data. The ANN model is simply a computational tool which needs to be combined with a systematic theoretical framework/strategy so that a time-consuming molecular simulation can be replaced without incurring significant loss of accuracy. While the role of the ANN model is to act as a computationally-inexpensive surrogate model, it is the framework that provides the basis for such calculations. Recently, ML has been applied to modelling of the free energy[10] and for thermodynamic calculations[11–13]. In this work, we discuss a different strategy for thermodynamic calculations which involves the combination of reverse Monte Carlo (RMC) and ANN models. The ANN model is used to solve a detailed balance equation.

The problem is of general interest to the area of statistical thermodynamics. Free energies associated with ideal gas, translation, rotation, vibration, electronic and nuclear degrees of freedom can be evaluated with reasonable accuracy. This leaves the challenging part, namely, the configurational free energy, which arises from the ordering/arrangement of molecules within the system[14–21]. A popular approach is the use of Metropolis Monte Carlo (MC) simulations[22–25]. In general, many configurations need to be sampled in MC before the calculation is converged. Thus, MC is computationally expensive and usually only a limited number of conditions can be simulated. MC calculations also become computationally prohibitive as the system size increases. The use of free energy extrapolations/expansions can help improve the efficiency[26–28]. In this paper, the configurational free energy is evaluated using the combination of RMC and ANN model, which completely bypasses the need for MC without compromising on accuracy.

The RMC-based framework for thermodynamic calculations [29–31] was recently introduced by the authors. We study adsorption on 2D lattices shown in Figure 1a and b. Such lattices are typically encountered with the face centered cubic (100) and (111) surfaces of metal catalysts. It should be noted that several machine learning models[32–38] have been developed to study adsorption phenomena on more complex systems with machine learning algorithms that are more sophisticated than the ANN model used here. Nonetheless, our approach extends the application of RMC to adsorption studies. This is because traditionally RMC has been mainly employed to create 2D/3D structural models for liquids and solids that are consistent with experimental scattering data [39–47], and not for thermodynamic calculations. The combined RMC-ANN modelling approach is shown to be accurate and orders-of-magnitude faster than MC. MATLAB codes with compact ANN models are provided in Supporting Information to demonstrate how configurational terms are evaluated in few seconds on a desktop computer.

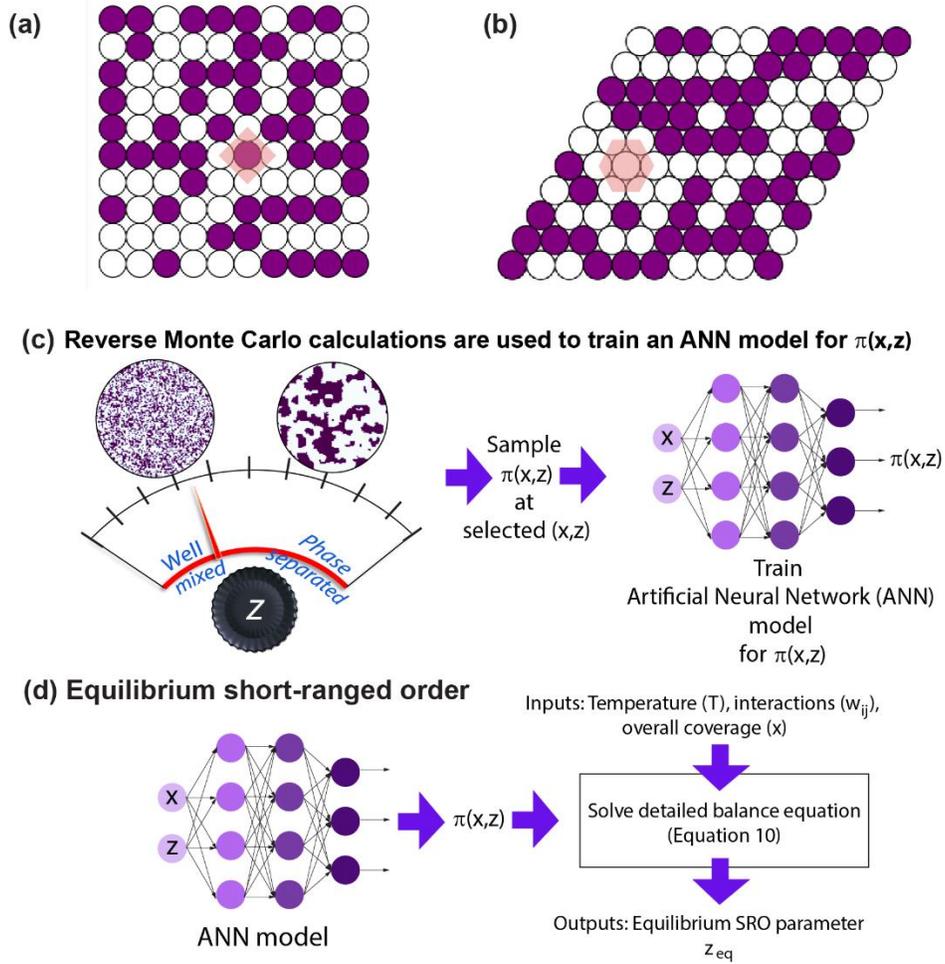

Figure 1: (a) Square and (b) triangular lattices. Filled/open circles denotes occupied/vacant sites. (c) Flowchart for the construction of ANN model for $\pi(x, z)$. (d) Flowchart for determining short-ranged order $z_{eq}$ at equilibrium.

The basic theory is developed in Section 2. In Section 3, we discuss the RMC algorithm, construction procedure of ANN, as well as GCMC calculations. In Section 4, we assess the accuracy of our ANN model. We validate our ANN-RMC approach by generating adsorption isotherms and making comparisons to grand-canonical Monte-Carlo (GCMC). Isotherm for the H/Ni(100) system is generated. Conclusions are provided in Section 5.

**2. Theory**

Consider a 2D lattice system denoted as $A_x V_{1-x}$, where $A$ is the adsorbed species, $V$ represents vacant sites and $x$ is the adsorbate coverage. $N_A, N_V$ and $N_t$ are the number of occupied, vacant, and total sites, respectively, $N_t = N_A + N_V$ and $x = N_A/N_t$. Adsorbed $A$ particles interact with first nearest neighbors (1NN) interactions. $w$ is the interaction energy for an $A - A$ pair. Periodic boundary conditions are employed. The Hamiltonian for the system is

$$H(\sigma; N_A, N_t, T) = \sum_{<i,j>} w\, \sigma_i \sigma_j. \qquad (1)$$

$\sigma_i$ is the occupation at site $i$. $\sigma_i$ is 1 when the site is occupied, and zero otherwise. $<i,j>$ in Equation (1) implies pairs of 1NN sites $i - j$.

**2.1 Chemical potential calculation using environment distribution**

Different local particle arrangements are possible (see Figure 1). $\epsilon$ denotes the number of $A$ atoms in the first coordination shell. The probability associated with such $\epsilon$ when the central site is occupied by $A$ is $\pi_{AA,eq}(\epsilon)$.[a] The subscript $eq$ implies equilibrium when $N_A$, $N_t$ and $T$ are constant. Similarly, $\pi_{VA,eq}(\epsilon)$ is the distribution around a $V$ site, such that $\pi_{AA,eq}(\epsilon) \neq \pi_{VA,eq}(\epsilon)$.

From Ref. [30], the chemical potential is

$$\Delta\mu(x,T) = k_B T \ln\frac{x}{1-x} - k_B T \ln\langle\exp(-\beta\Delta U_\epsilon)\rangle_x \qquad (2)$$

where $\beta = (k_B T)^{-1}$, $k_B$ is the Boltzmann constant, $T$ is the temperature, and

$$\langle\exp(-\beta\Delta U_\epsilon)\rangle_x = \sum_{\epsilon=0}^{c} \pi_{VA,eq}(\epsilon)\exp(-\beta\Delta U_\epsilon). \qquad (3)$$

The coordination number $c$ in the 1NN shell is 4 for square lattice and 6 for triangular lattice. $\Delta U_\epsilon$ is the energy change when an $A$ particle is inserted at a vacant site, i.e., $\Delta U_\epsilon = \epsilon w$. Alternatively, $\Delta\mu$ is calculated by removing an adsorbed $A$, i.e.,

$$\Delta\mu(x,T) = k_B T \ln\frac{x}{1-x} + k_B T \ln\langle\exp(-\beta\Delta U'_\epsilon)\rangle_x. \qquad (4)$$

In such a case, the excess term

$$\langle\exp(-\beta\Delta U'_\epsilon)\rangle_x = \sum_{\epsilon=0}^{c} \pi_{AA,eq}(\epsilon)\exp(-\beta\Delta U'_\epsilon). \qquad (5)$$

Here, $\Delta U'_\epsilon$ is the energy change associated with the removal of an $A$ particle with environment $\epsilon$ from an occupied site.

## 2.2 Local environment, probability distribution and connection to short-ranged order

From equation (2)-(5), we find that $\pi_{AA,eq}(\epsilon)$ and $\pi_{VA,eq}(\epsilon)$ are key quantities for estimating thermodynamic properties. A procedure to calculate $\pi_{AA,eq}(\epsilon)$ and $\pi_{VA,eq}(\epsilon)$ is needed. We assume that the 2D atomic arrangements at the surface are determined by $x$ and a short-ranged order (SRO) parameter $z$[39,48–50]. $z$ is the average fraction of 1NN sites around $A$ occupied by $A$ particles.[29] $z$ determines the local ordering. See Figure 1c for types of arrangements associated with different values of $z$. For random arrangement of the adsorbed particles $z = x$. Separate $A$- and $B$-rich regions are formed when $z \approx 1$.

---

[a] In general, $\pi_{ij}(\epsilon)$ is the probability of having $\epsilon$ number of $j$ atoms around central site with $i$ atom.

How $\pi_{AA}$ and $\pi_{VA}$ varies with $x$ and $z$ is ascertained using RMC. $x$ and $z$ are the main inputs to RMC. Initially, a 2D lattice configuration random $A$ arrangement is created. Trial moves are attempted wherein the positions of a randomly chosen pair of $A$ and $V$ sites are swapped with an acceptance probability. The procedure is repeated till the target number of $A - A$ first nearest neighbor pairs $N_{AA,target} = \frac{czN_A}{2}$ is achieved. RMC calculations are independent of $w$ and $T$.

2D RMC configurations obtained at selected values of $(x, z)$ are used to evaluate $\pi_{AA}(\epsilon; x, z)$ and $\pi_{VA}(\epsilon; x, z)$. Figure 1c shows a flowchart for the construction of the ANN model for $\pi(\epsilon; x, z)$. The ANN model provides a closed form expression for $\pi(\epsilon; x, z)$, which is not generally available for 2D lattice systems.

### 2.3 Environment distribution at equilibrium using the Detailed Balance Equation

We write $\pi_{AA,eq}(\epsilon) \equiv \pi_{AA}(\epsilon; x, z_{eq})$ and $\pi_{VA,eq}(\epsilon) \equiv \pi_{VA}(\epsilon; x, z_{eq})$. This implies that the equilibrium distribution is the one obtained from RMC at $(x, z_{eq})$. The goal is to determine $z_{eq}$, the SRO parameter value at equilibrium, by solving a detailed balance equation.

Consider a swap move $A(\epsilon) + V(\epsilon') \rightleftharpoons A(\epsilon') + V(\epsilon)$, where $A(\epsilon)$ implies an $A$ with environment $\epsilon$. For $\epsilon < \epsilon'$, the forward direction (right arrow) results in an $A$ particle moving into an $A$-rich environment. The net probability flux for this combination $\epsilon - \epsilon'$ is

$$f(\epsilon, \epsilon'; x, z) = p(A(\epsilon), V(\epsilon'))\Gamma_r - p(A(\epsilon'), V(\epsilon))\Gamma_l. \quad (6)$$

$\Gamma_r$ and $\Gamma_l$ are transition probabilities in the right and left direction, respectively. The probability of selecting the pair $A(\epsilon) - V(\epsilon')$ is $p(A(\epsilon), V(\epsilon')) = \pi_{AA}(\epsilon)\pi_{VA}(\epsilon')$. Similarly, the probability of selecting the pair $A(\epsilon') - V(\epsilon)$ is $p(A(\epsilon'), V(\epsilon)) = \pi_{AA}(\epsilon')\pi_{VA}(\epsilon)$. The ratio $\frac{p(A(\epsilon'), V(\epsilon))}{p(A(\epsilon), V(\epsilon'))} = \exp(-\beta \Delta U)$ where $\Delta U = w(\epsilon' - \epsilon)$ is the energy change of the system along the right arrow. $\Gamma_r$ and $\Gamma_l$ are expressed analogous to the standard Metropolis acceptance criterion[51] as

$$\Gamma_r = \min(1, \exp(-\beta \Delta U)) \quad (7)$$

and

$$\Gamma_l = \min(1, \exp(\beta \Delta U)). \quad (8)$$

The overall net probability flux obtained by considering all pair environments $\epsilon - \epsilon'$

$$N(x, z) = \left| \sum_{\epsilon < \epsilon'} f(\epsilon, \epsilon'; x, z) \right|. \quad (9)$$

is zero at equilibrium. Finally, we write the detailed balance equation as

$$N(z_{eq}; x, w, T) = 0. \quad (10)$$

A crucial point is that $\pi(\epsilon; x, z)$ can be employed with a variety of interactions, temperatures and coverages. This makes the ANN model versatile. When $x$, total number of lattice sites $N_t$, temperature $T$ and the adsorbate-adsorbate interactions $w$ are specified, the goal is to identify the value of $z_{eq}(x,w,T)$ where detailed balance is satisfied. Equation (10) is solved using a gradient-based Newton method. The flowchart is shown in Figure 1d. The approach is computationally less demanding than MC since it involves root-finding. Thereafter, $\Delta\mu$ is calculated using Equation (2) or (4).

ANN offers some significant implementational advantages compared to our earlier approach. Earlier, $\pi(\epsilon;x,z)$ used to be stored in the form of look-up tables[29–31]. Look-up tables can be unwieldy, whereas ANN is compact. A Delaunay triangulation-based interpolation scheme was employed with the look-up table to calculate $\pi$ at any given $x, z$. A fine-grid in $x - z$ space[31] is needed to lower the interpolation error. ANN requires RMC structures at fewer number of $x - z$ points as shown later. Moreover, the interpolated probability from look-up tables can contain noise, which prohibits the use of gradient-based Newton method for root finding.

## 3. Computational details

### 3.1 RMC algorithm

A brief description of the RMC algorithm is provided. A periodic lattice of size $325 \times 325$ with $N_t = 105{,}625$ is employed. A random arrangement of $A$ is prepared, such that initially the number of $A - A$ bonds $N_{AA} = \frac{1}{2}N_A cx$. The steps in RMC are [29]:

Step 1: $N_{AA}$ is calculated for current configuration. The distance from the target structure $d^2 = \left|N_{AA} - N_{AA,target}\right|^2$ is calculated.

Step 2: A pair of $A$ and $V$ sites are chosen randomly and their positions are swapped. The new number of $A - A$ bonds $N_{AA,n}$ as well as the new distance to target structure $d_n^2$ is evaluated.

Step 3: The acceptance probability is calculated as

$$p_{acc} = \min(1, \exp(d^2 - d_n^2)). \qquad (11)$$

The lattice structure is updated if the move is accepted. Else, the old configuration remains.

Step 4: The RMC calculation is stopped once $\pi_{AA}$ and $\pi_{VA}$ become stationary. Otherwise, steps 1-4 are repeated.

Snapshots from RMC are shown in Figure 2. Usually, ordering behavior is expected when $z < x$. Random structures are obtained with $x = z$. Clustering of $A$ is observed with increasing $z$.

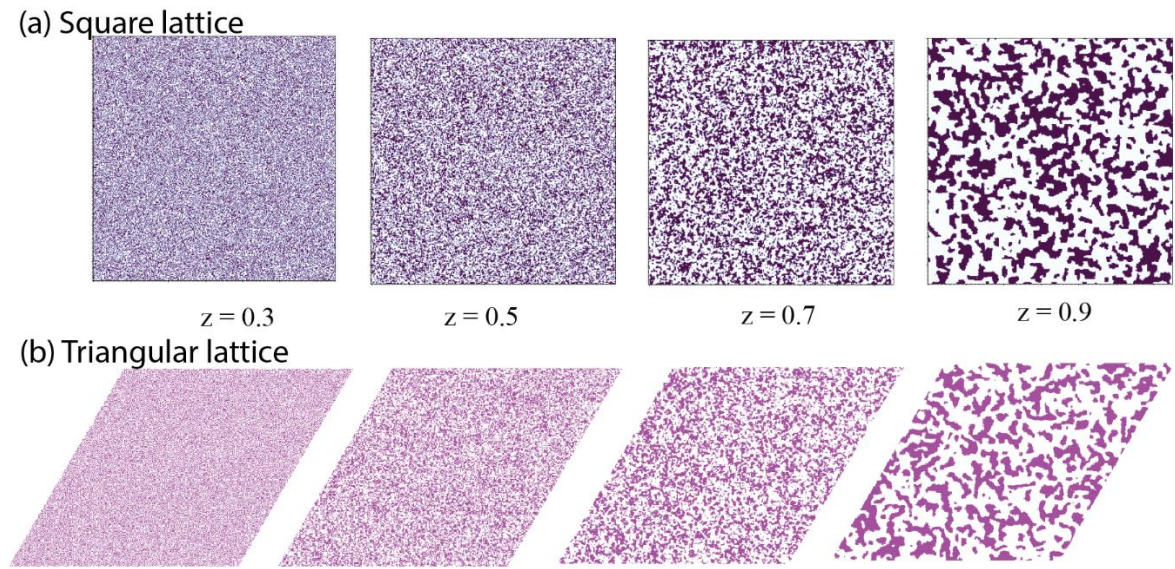

Figure 2: RMC structures obtained with $x = 0.4$ and varying $z$. (a) Square and (b) triangular lattice. Purple dots indicate $A$-sites.

## 3.2 ANN construction

The goal here is to "learn" a function that maps $x$ and $z$ to probability distributions discussed earlier. The function is inferred using the data $\pi_{AA}(\epsilon; x, z)$ and $\pi_{VA}(\epsilon; x, z)$, $\epsilon \in [0, c]$ collected using RMC. A large number of supervised learning methods are available[52]. However, because of our previous experience with ANN models[9,53], we have preferred ANN. Figure 3a shows the architecture used to obtain a probability distribution. The inputs to the ANNs are $x$ and $z$. There are $c + 1$ different $\epsilon$ values. One ANN can predict $\pi_{AA}(\epsilon; x, z)$ for a given $\epsilon$. Thus, $c + 1$ ANNs are used. The situation is similar for $\pi_{VA}(\epsilon)$. Finally, $\pi_{AA}(\epsilon)$ or $\pi_{VA}(\epsilon)$ is normalized so that $\sum_{\epsilon=0}^{c} \pi_{AA}(\epsilon) = 1$ and $\sum_{\epsilon=0}^{c} \pi_{BA}(\epsilon) = 1$.

375 points sampled in the $x - z$ space for training and validation are shown in Figure 3b. Such a dataset was not available previously. RMC calculations were performed for the square and triangular lattice separately. Although in principle $x, z \in [0,1]$, certain parts of the space are not accessible with RMC. For instance, the values of $z < 1.2636x - 0.263$ are not topologically possible especially when $x > 0.2$, as this would require low number of $A$ pairs despite the high fraction of $A$ particles in the lattice. This approximate boundary is shown as a solid blue line in Figure 3b. Similarly, configurations with $z$ values above the dashed blue line in Figure 3b cannot be created with the size of the lattice chosen in the present study especially when $x$ is low, as it will require almost every $A$ to be surrounded by $A$ neighbor although the fraction of $V$ sites is large. The inaccessible region is approximately $z > 0.334\,x^3 - 0.6197\,x^2 + 0.4039x + 0.882$. This boundary can be shifted towards higher $z$ by selecting a lattice of larger size.

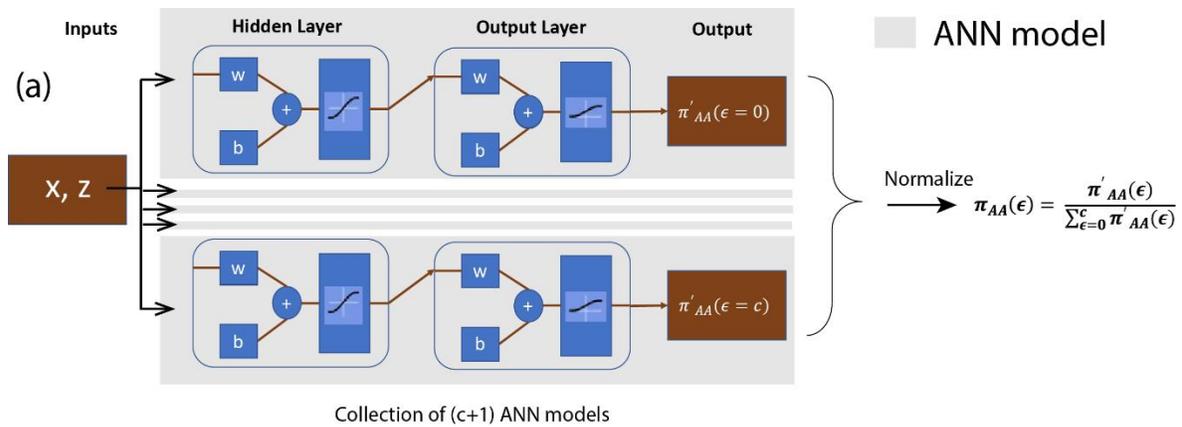

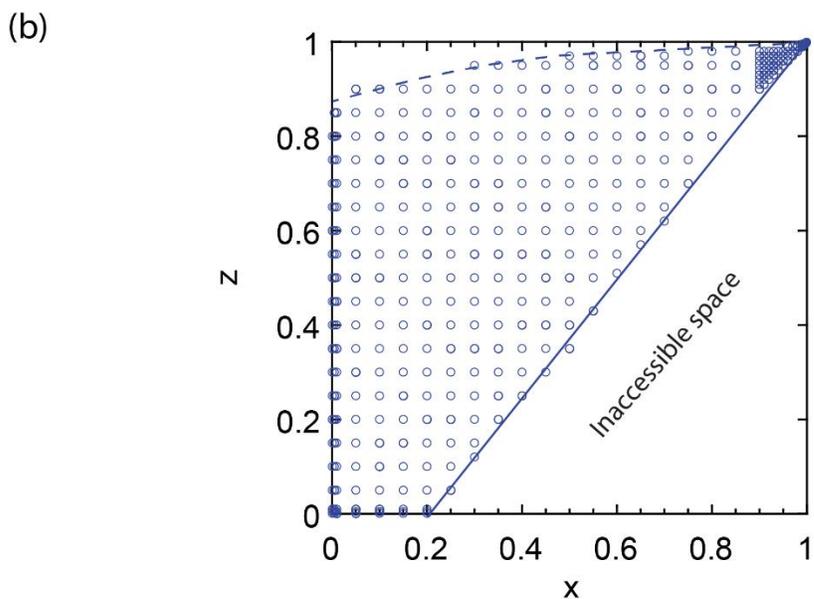

Figure 3: (a) ANN model structure: each model gives the probability for a particular $\epsilon, \epsilon \in [0, c]$. The output is normalized so that sum of probabilities is one. (b) $x - z$ space sampled using RMC. Each circle denotes one RMC calculation.

The ANN training is performed using feedforward neural network with 1 to 2 hidden layers in MATLAB (see Section S1 of Supporting Information). The number of neurons in the hidden layer(s) is allowed to vary between 5 to 10. We begin with a single neuron and then keep adding neurons until the accuracy of training and validation increases. Beyond a certain point, the accuracy of validation decreases with additional neurons due to overfitting. At this point, we obtain the optimal number of neurons. Hyperbolic tangent (tanh) is used as the activation function in the hidden layer(s) due to its differentiability over the entire domain. Sigmoid function is used in the output layer so that the probabilities lie between 0 and 1. The probability distribution dataset which is used to train and validate ANN models does not require additional scaling as it is already normalized. Mean Squared Error (MSE) is used as loss function while training the ANN models. The weights and biases of ANNs are initialized using Nguyen-Widrow initialization algorithm[54].

The training set is generated by random selection of 70 % data of the original dataset, the rest is incorporated in the validation set. The dataset consists of 375 datapoints. This is far smaller than earlier datasets used with interpolation method – such datasets contained >800 points[31]. A Levenberg-Marquardt backpropagation algorithm is used for minimizing the mean-squared error (MSE) between predicted and actual probability. The number of epochs is set to be $10^5$ and the tolerance for gradient in MSE is $10^{-7}$. Section S1 of Supporting Information provides details of the learning curve, MAE, MSE reduction behavior and cross-validation.

**3.3 GCMC calculations**

Grand-canonical ensemble (constant $\Delta\mu$, $T$, $N_T$) MC calculations are performed to determine $x$ as a function of $\beta\Delta\mu$ for different interactions $w$. Two types of starting configurations are considered: only $V$ (loading) and only $A$ present (unloading). The GCMC calculation consists of swap, insertion and deletion trial moves.[55] Acceptance criterion is shown in Table 1.

Table 1: Probability acceptance criterion of different trial moves. $\Delta U$ denotes the energy difference between the new and the old configuration.

| Type of trial move | Percentage of moves attempted | Acceptance probability ($p_{acc}$) |
|---|---|---|
| Swap | 60 | $\min(1, \exp(-\beta \Delta U))$ |
| Insertion | 20 | $\min\left(1, \dfrac{N_B}{N_A + 1} \exp[\beta(\Delta\mu - \Delta U)]\right)$ |
| Deletion | 20 | $\min\left(1, \dfrac{N_A}{N_B + 1} \exp[-\beta(\Delta\mu + \Delta U)]\right)$ |

## 4. Results and discussions

### 4.1 Accuracy of the ANN model

The parity plots in Figure 4 show the ANN predicted probabilities versus probabilities from RMC with the validation dataset. The training and validation results are shown separately in Section S1 of Supporting Information. Excellent agreement is observed for both square and triangular lattice. The values of correlation coefficient $R$ during training and validation is presented in Table 2. Large values of $R$ exceeding 0.999 are obtained. This confirms that $\pi_{AA}(\epsilon; x, z)$ and $\pi_{VA}(\epsilon; x, z)$ are accurately captured by the ANNs in the entire $x - z$ space of interest.

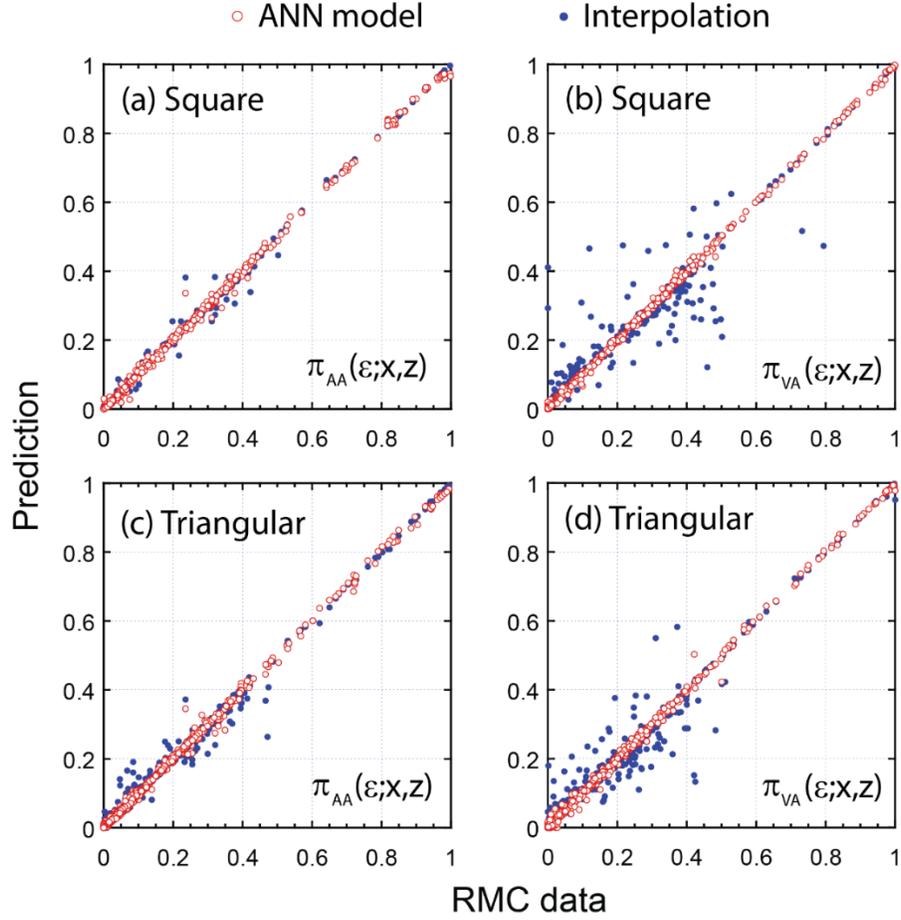

Figure 4: Parity plots showing predicted versus actual probabilities obtained with the validation data. Red open circle and blue filled circles denote ANN and interpolation methods, respectively.

Table 2: R values during training and validation of the ANNs

| Lattice | Probability distribution | Training step $R$ | Validation step $R$ | Overall R |
|---|---|---|---|---|
| Square | $\pi_{AA}$ | 0.9993 | 0.9994 | 0.9993 |
| | $\pi_{VA}$ | 0.9997 | 0.9997 | 0.9997 |
| Triangular | $\pi_{AA}$ | 0.9992 | 0.9993 | 0.9992 |
|  | $\pi_{VA}$ | 0.9996 | 0.9994 | 0.9995 |

In general, a large number of data points are required to train an ANN model. However, this issue is not encountered here because the $\pi_{AA}(\epsilon; x, z)$ and $\pi_{VA}(\epsilon; x, z)$ functions do not possess significant undulation. Therefore, it is possible to create a compact ANN model that is reasonably accurate. A comparison of ANN and interpolation scheme shows that the former is superior. Both approaches used the same training data set to predict the probabilities in the validation dataset. Figure 4 shows that the ANN model can achieve reasonable accuracy with fewer data points than the interpolation scheme. The maximum absolute error with the interpolation method using the validation dataset are 0.15 and 0.41 for $\pi_{AA}$ and $\pi_{VA}$ of square lattice, respectively. Corresponding error for triangular lattice is 0.21 and 0.3.

$\pi_{AA}$ and $\pi_{VA}$ from the ANN models are shown as barplots in Figures 5 and 6. $\pi_{AA}$ and $\pi_{VA}$ for a given $\epsilon$ can be non-monotonic. For example, see variation in $\pi_{AA}(\epsilon = 2; x = 0.4, z)$ with respect to $z$ in Figure 5. Once again we can see that overall the probabilities are correctly predicted, i.e., the absolute error is small. However, the percentage error is large when the probabilities are small. This is expected since the absolute error is used in the training procedure. Environments associated with small probabilities are more susceptible to sampling errors in RMC since very large lattices or multipe samples are required. The error in Equation (10) due to such $\pi_{AA}$ and $\pi_{VA}$ is not amplified, as small probabilities do not contribute significantly.

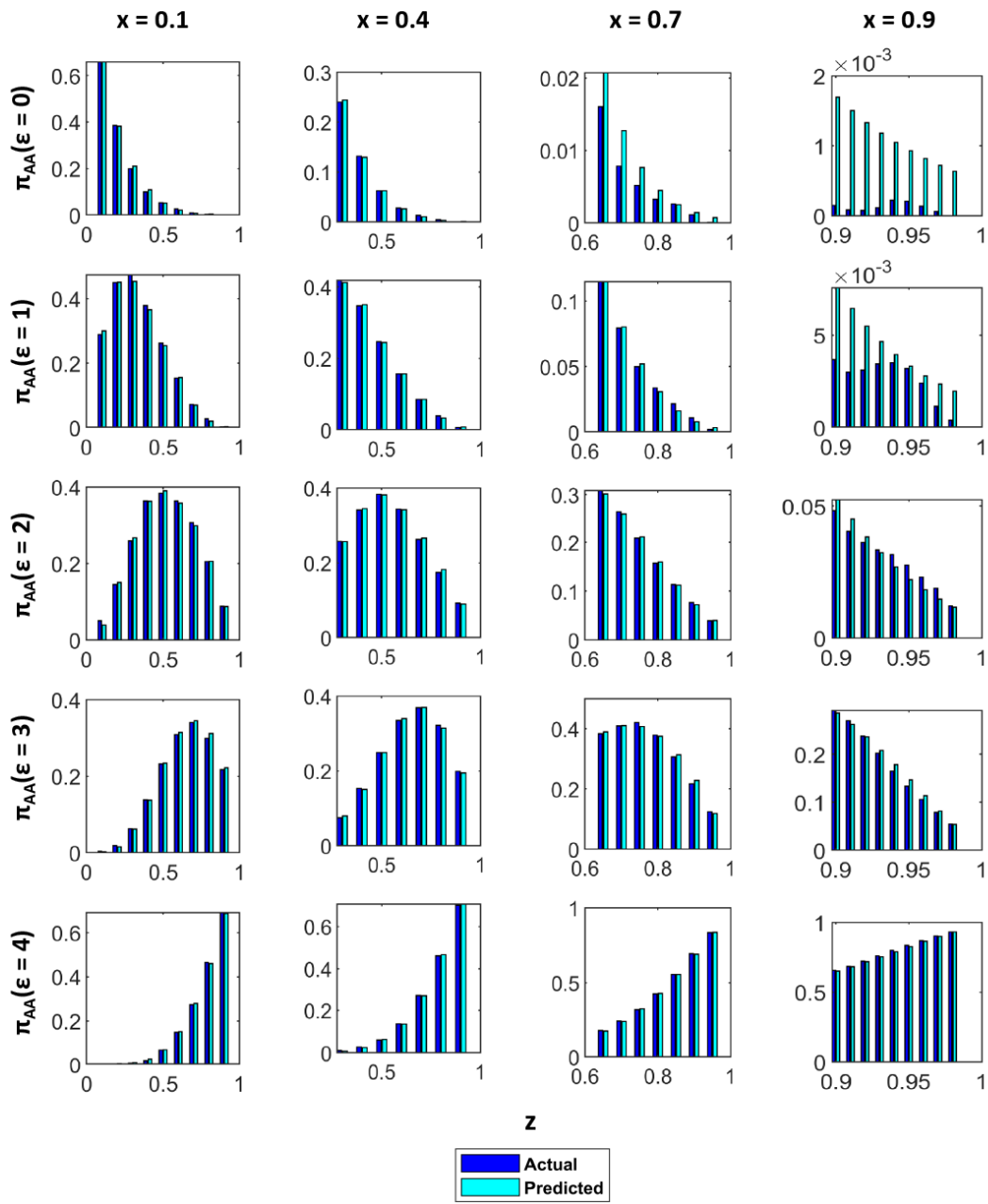

Figure 5: Barplot showing actual and ANN predicted $\pi_{AA}(\epsilon)$ for square lattice.

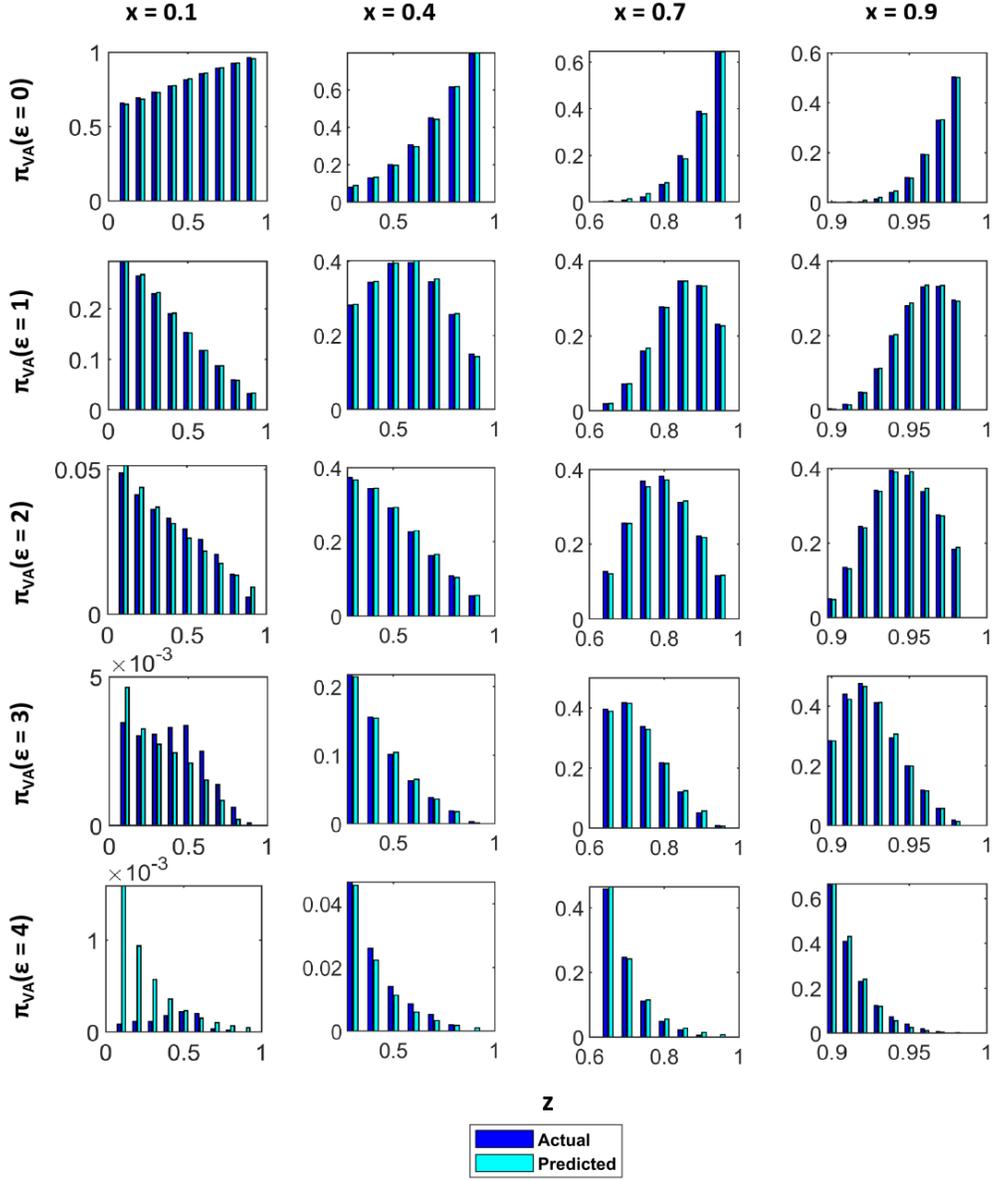

Figure 6: Barplot showing actual and ANN predicted $\pi_{VA}(\epsilon)$ for square lattice.

The variation in probability distributions depends on the region of $x - z$ space under consideration. Therefore, a variable grid size in $x - z$ space is used in Figure 3b. Large variations are observed with dilute/dense systems where $x$ and $z$ take extreme values ($x, z \geq 0.9$). A grid size as low as 0.001 is used at some locations. A relatively coarser grid of gridsize 0.05 is used for other regions of $x - z$ space.

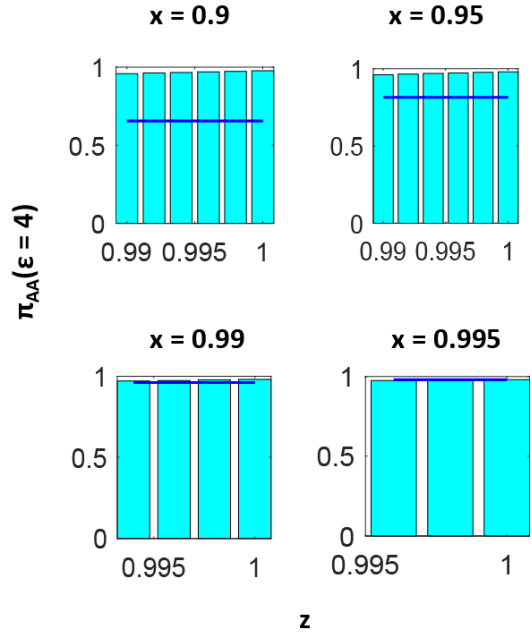

Figure 7: $\pi_{AA}$ predicted by ANN models in parts of the $x - z$ space for square lattice where RMC calculations do not converge. The blue lines respresent probabilities from the binomial distribution.

Recall the region above the dashed line in Figure 3b which is inaccessible to RMC. We explore the ANN behavior when it is extrapolated to $z = 1$ where $x \geq 0.9$. This corresponds to the top-right corner of the $x - z$ space in Figure 3b. In Figure 7, the ANN predicted $\pi_{AA}$ for square lattice is shown in terms of barplots. The horizontal axis corresponds to $z$. Only large values of $x$ ($x = 0.9, 0.95, 0.99$ and $0.995$) are considered in the panels and $\epsilon = 4$. The largest value of $z$ that were sampled with these $x$ are 0.98, 0.98, 0.993 and 0.995, respectively. From Figure 7 we see that the ANN model is well-behaved till $z = 1$. Based on these results, we can demarcate the top inaccessible part of $x - z$ space in Figure 3b as $z > 0.334 \, x^3 - 0.6197 \, x^2 + 0.4039x + 0.882$ provided $x < 0.9$. For $x > 0.9$, the ANN model can be used till $z = 1$.

The binomial distribution is generally valid for dilute systems. Blue lines in Figure 7 represent the probabilites based on the binomial distribution $\pi_{AA}^{binomial}(\epsilon) = \frac{c!}{\epsilon!(c-\epsilon)!} x^\epsilon (1-x)^{c-\epsilon}, \epsilon \in [0, c]$. At $x = 0.9$, the ANN predicted $\pi_{AA}(\epsilon = 4)$ is quite different from the binomial distribution. This is because the $z$ values are much larger than $x$. Therefore, $x = 0.9$ cannot be considered as an ideal system unless $z \approx x$. As $x$ increases, the predicted values approach the binomial distribution for any $z > x$. Good agreement between the ANN model and binomial distribution is observed for $x = 0.995$. Thus, the system can considered to be ideal for $x \geq 0.995$.

Similar to square lattice, we provide barplots for the actual and ANN predicted probabilities in Figures 8 and 9. Once again, probabilities are correctly predicted. A comparison between Figures 5 and 6 with Figures 8 and 9 reveals that the trends for the square and triangular lattice are quite similar. The value of the probabilities are observed with the triangular lattice are slightly lower since more number of $\epsilon$ values are possible. For example, when $x = 0.1$ the probability $\pi_{AA}(\epsilon = 0, z = 0.1)$ is 0.62 and 0.5 in case of square and triangular lattices, respectively.

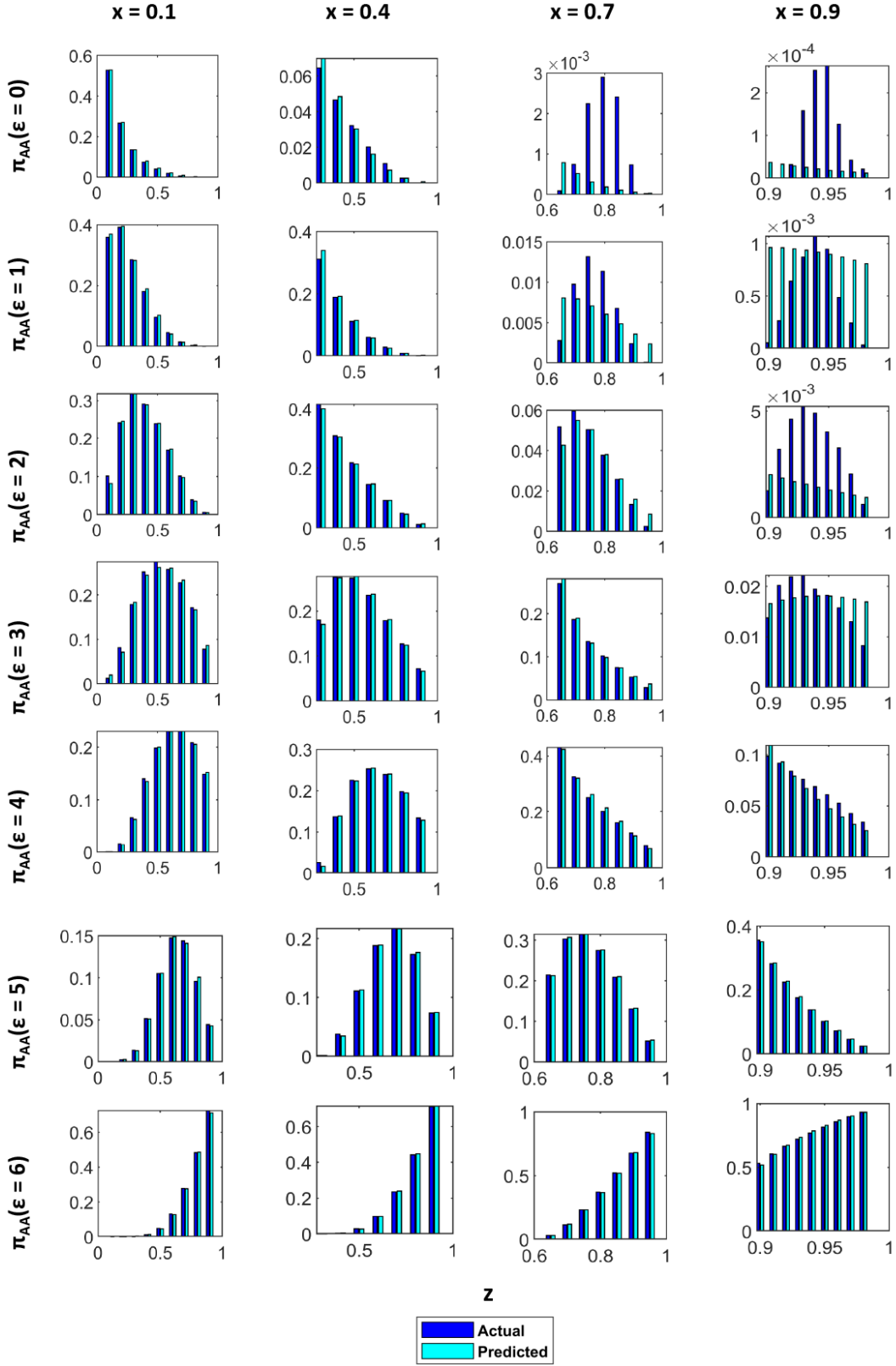

Figure 8: Barplot showing actual and ANN predicted $\pi_{AA}(\epsilon)$ for triangular lattice.

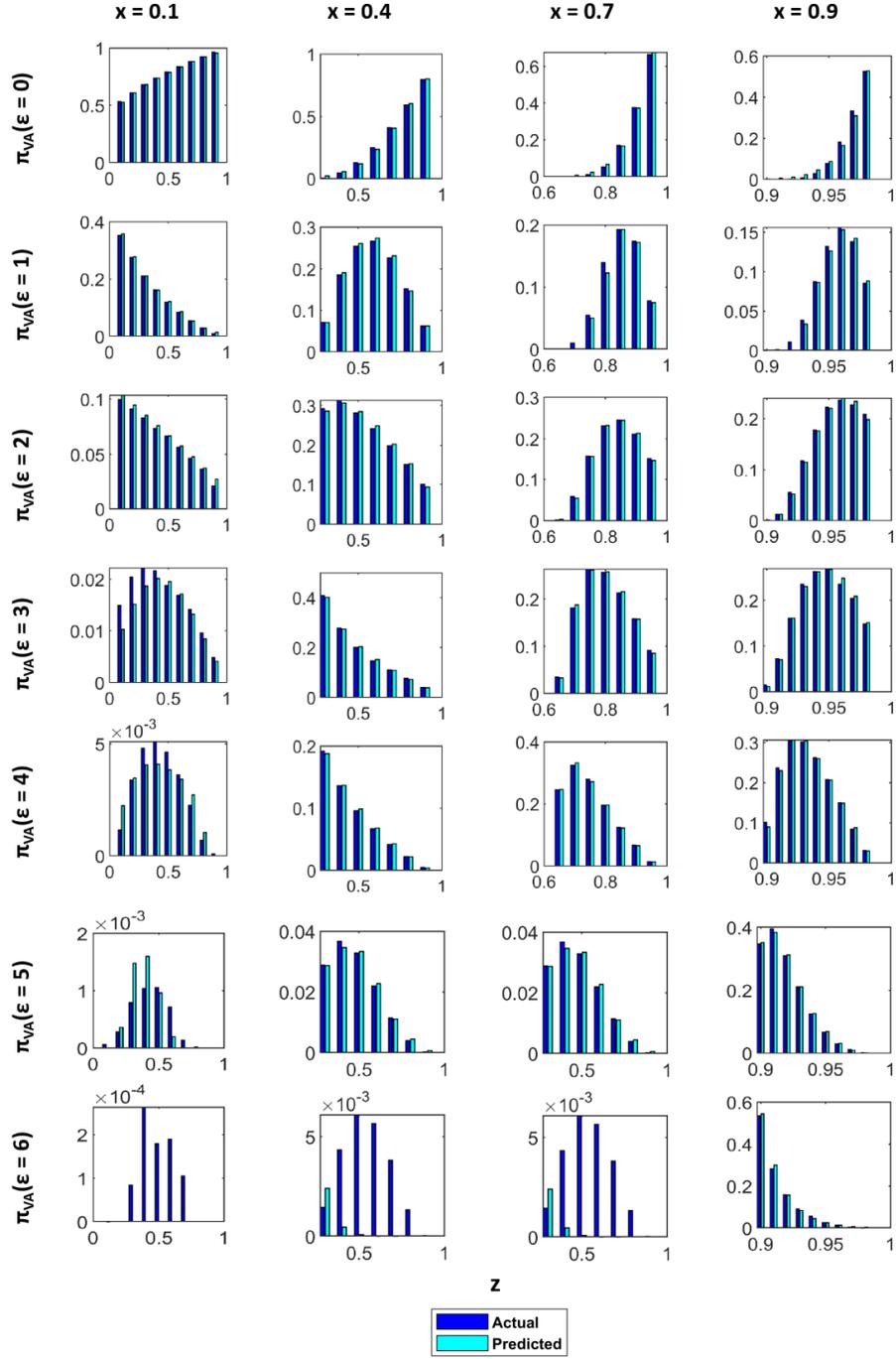

Figure 9: Barplot showing actual and ANN predicted $\pi_{VA}(\epsilon)$ for triangular lattice.

## 4.2 Net probability flux

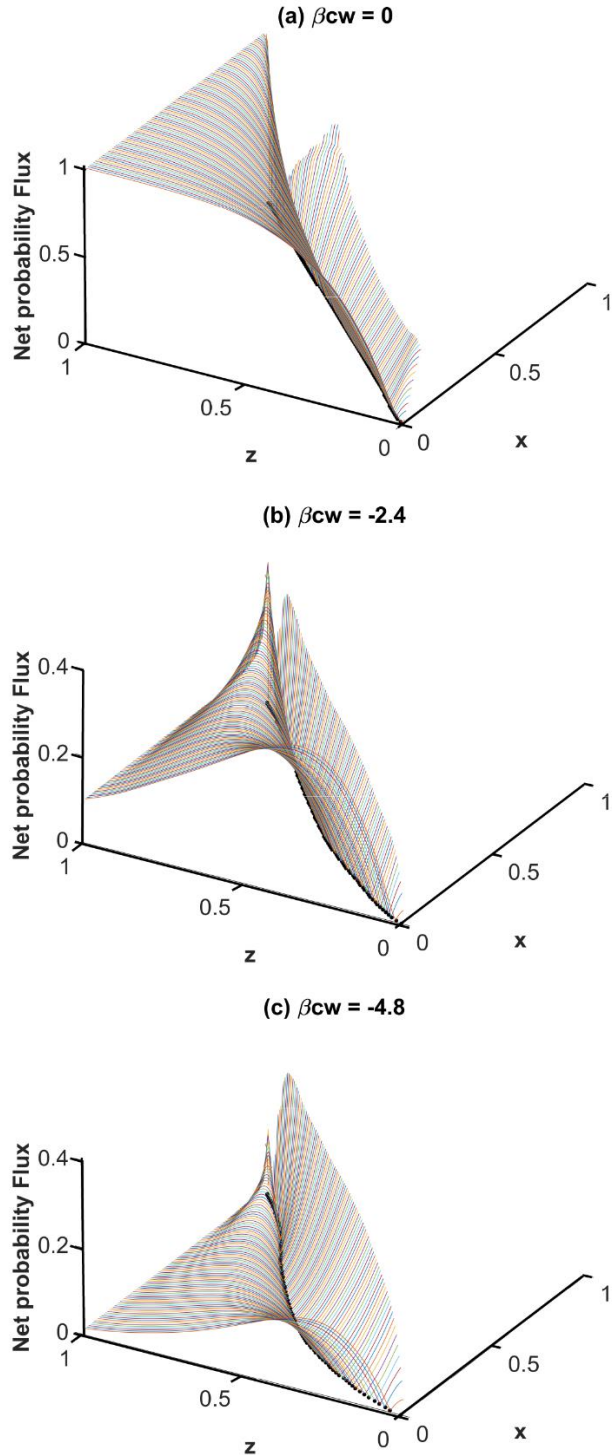

Figure 10: Net probability flux calculated at different temperatures/interactions for square lattice. Black circles denote $z_{eq}$ where the net flux becomes zero.

To illustrate how the equilibrium short-ranged order parameter $z_{eq}$ is evaluated, we use the ANN model for $\pi_{AA}$ and $\pi_{VA}$ with Equation (9) to calculate the net probability flux as a function of $x$ and $z$. The result is presented as 3D line plot in Figure 10 for different $\beta cw$. Black circles are the points where the net flux becomes zero, i.e., it yields the $z_{eq}(x, w, T)$. When interactions are absent (panel a), i.e., $\beta cw = 0$, we find $z_{eq} = x$. This is in agreement with our expectation of a perfectly random arrangement of $A$. For stronger attractive interactions $\beta cw < 0$, $z_{eq}$ is always greater than $x$ implying clustering of $A$ will be observed. Similar results are obtained for triangular lattice (not shown). A MATLAB code illustrating the calculation of $z_{eq}$ and $\Delta\mu$ with square/triangular lattice has been provided in Section S2 of the Supporting Information. Figure 11 shows how $z_{eq}$ varies with $x$ for both lattices. For a given $x$ and $\beta cw$, $z_{eq}$ obtained for the triangular lattice is slightly lower than the one for square lattice.

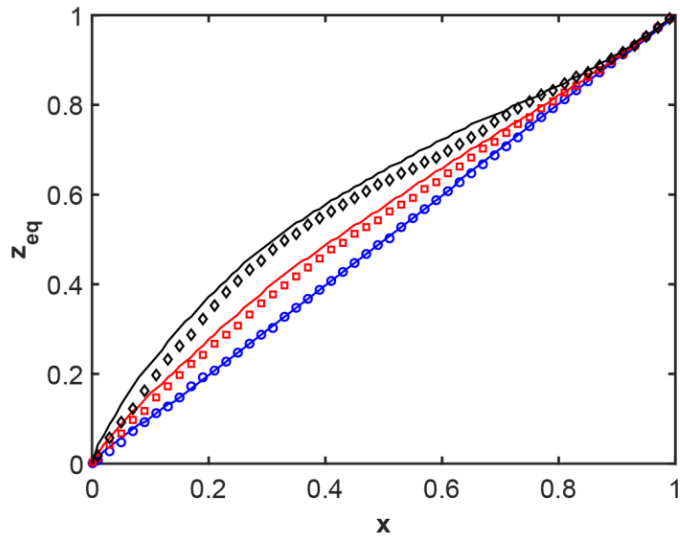

Figure 11: $z_{eq}$ as a function of $x$. Lines represent square lattice, whereas symbols represent triangular lattice. Blue: $\beta cw = 0$, Red: $\beta cw = -2.4$, Black: $\beta cw = -4.8$.

## 4.3 Adsorption isotherm

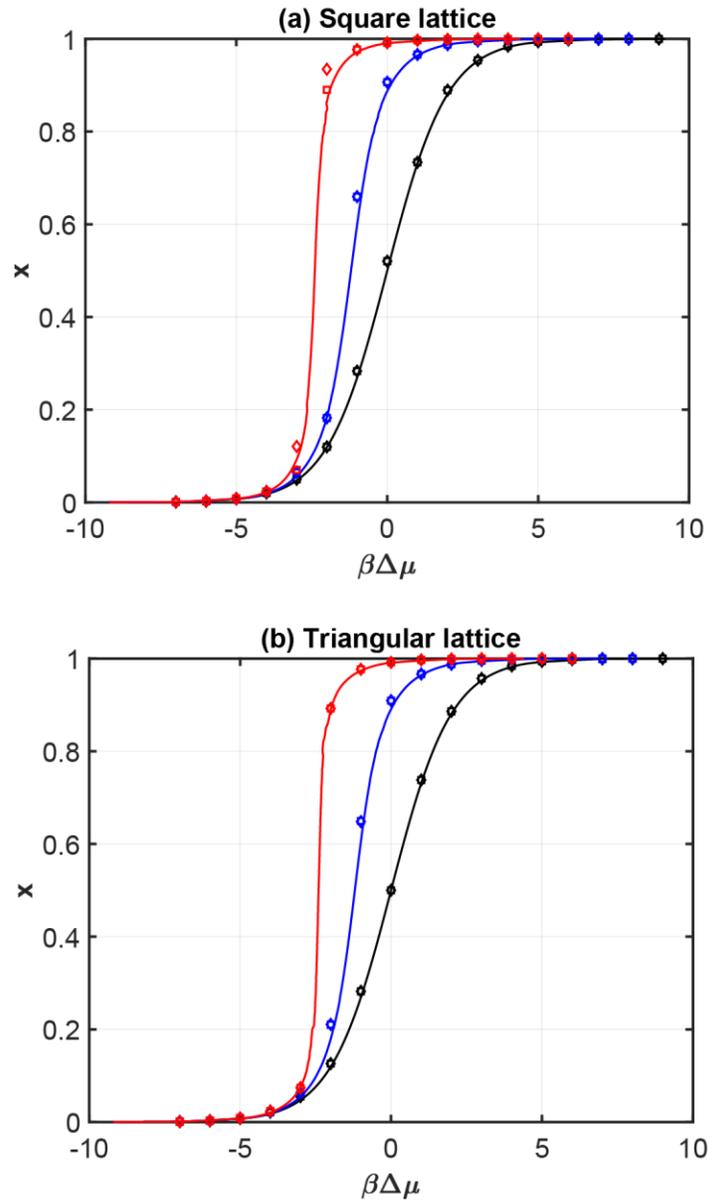

Figure 12: Chemical potential estimation using ANN/RMC (lines) and GCMC (symbols) with varying $\beta cw$. Black: $\beta cw = 0$, Blue: $\beta cw = -2.4$, Red: $\beta cw = -4.8$. Square symbols represents GCMC loading calculations, whereas diamond symbols represents unloading.

Transferability and extendibility are two important aspects related to the use of ML models in molecular simulations. A common application involves the prediction of atom-specific information, such as partial charges[56], force fields[57], etc. In these applications, transferability refers to the use of the model with any molecular system as long as the properties of constituent atoms can be predicted. Transferability ensures data efficiency, low computational overhead and breadth of application. In the present work, the ANN models for statistical distributions like $\pi_{AA}$ and $\pi_{VA}$ are transferable in the sense that they can apply to a number of atomic/molecular systems and a wide range of interactions strengths and temperatures as shown next. Extendibility refers to the use of the model with a system of size larger than the one it was first trained with. The ANN models developed here are extendable. Properties such as energy, free energy changes and number of $A - A$ pairs can be easily estimated for larger system sizes with the same ANN model.

Figure 12 shows the adsorption isotherms generated using ANN/RMC model for $\beta cw$ between 0 to -4.8. We compare the isotherms with the ones generated using GCMC. In GCMC simulations, one calculates average $x$ at given value of $\beta \Delta \mu$ and $\beta cw$. Results for both types of starting structure are shown: pure $B$ (loading) and pure $A$ (unloading). We observe excellent agreement between the two approaches. The chemical potential $\Delta\mu = \frac{cw}{2}$ at $x = 0.5$ for both square and triangular lattice.

In most systems, the adsorbate-adsorbate interactions $w$ are fixed, whereas the temperature is varied. The values of $\beta cw$ considered in Figure 12 are in accordance with real systems, e.g., O adsorption on Pd (100) and Pt (111) surfaces. For (100) surface, adsorbed O occupy four-fold hollow sites[58] which corresponds to a square lattice. Similarly, O occupies the hollow FCC sites at the (111) surface[59] which corresponds to a triangular lattice. Considering the interactions in O-Pt (111) and O-Pd (100), the range of $\beta cw$ considered here corresponds to 483 K or higher, respectively. Typical experimental temperatures can be higher.

**4.4 Extension to off-lattice systems: H adsorption on Ni (100) surface**

Our ANN/RMC approach can be extended to off-lattice system. For this purpose, the surface adsorption for hydrogen on Ni(100) surface is studied. The adsorption isotherm from ANN/RMC and GCMC simulations are compared. The off-lattice GCMC model uses embedded atom method (EAM) potential for the Ni-H system[60]. The EAM interatomic potential provides the interactions between Ni-Ni, H-H and Ni-H atoms. The total potential energy of the system is given by

$$E(r_1, r_2, \ldots, r_N) = \sum_{\langle i,j \rangle} \phi_{ij}(r_{ij}) + \sum_i F_i(\rho_i). \qquad (12)$$

Here $r_i$ is the coordinate of atom $i$, $\phi_{ij}$ denotes the pair potential for $i$ and $j$ ($\phi_{H-H}$, $\phi_{Ni-H}$ or $\phi_{Ni-Ni}$), $r_{ij}$ is the separation between atoms $i$ and $j$, $F_i$ is the embedding energy for atom $i$, and $\rho_i$ is the density term. $\rho_i$ contains contributions from neighbors of atom $i$, which makes the EAM a many-body potential.

The LAMMPS code[61] is used to perform the iterative GCMC-MD calculations wherein 1500 GCMC trial moves is followed by 100 molecular dynamics (MD) steps. A total of 9 million trial moves are attempted in the GCMC calculations. The chemical potential of H is specified. The GCMC trial moves consisting of insertion of H, deletion of H and displacement of H are performed with an equal probability. The MD calculation ensures that the overall structure is relaxed. The average H coverage is obtained once the system has equilabrated. From GCMC calculations, we find that H occupies the four-fold hollow site. See the square lattice arrangement at low and high coverages in Figure 13a. It is assumed that the four-fold hollow site site is involved with ANN/RMC. Only up to 0.5 ML coverge is studied with GCMC, since two types of sites are found to be occupied at higher coverages[62].

The EAM interaction cutoff is beyond first nearest neighbor. However, the ANN/RMC framework we have presented employs the 1NN interaction strength $w$. We exploit the fact that the H-H and H-Ni interaction strength decays fairly quickly with respect to the distance between neighbors. The goal is to obtain an effective interaction parameter $w$ for the H-Ni system. A cluster expansion model[63] is constructed to estimate the configurational energy for the ANN/RMC framework:

$$E(\sigma) = \overbrace{\sum_i V_i \sigma_i}^{1-body} + \overbrace{\sum_{i>j} V_{i,j} \sigma_i \sigma_j}^{2-body} + \overbrace{\sum_{i>j>k} V_{i,j,k} \sigma_i \sigma_j \sigma_k}^{3-body} + \cdots. \tag{13}$$

$\sigma$ is the occupation vector and for any site $i$, $\sigma_i = 1$ if the site is occupied by H, otherwise $\sigma_i = 0$. The coefficients in the equation ($Vs$) are the effective cluster interactions (ECI) and they represent contribution from each cluster (1-body, 2-body etc.) to the energy. A large number of random H/Ni(100) configurations were generated. The energy associated with the energy-minimized configuration was obtained and the ECIs were fitted to such data. $V_i$ ($= -2.82\ eV$) corresponds to the binding energy of H. Since the interactions between the adsorbed H-H is weak, only 1 NN pair interactions ($V_{ij} = w = 0.0082\ eV$) are included. This allows us to use the ANN model from Section 4.1 for H/Ni(100) adsorption.

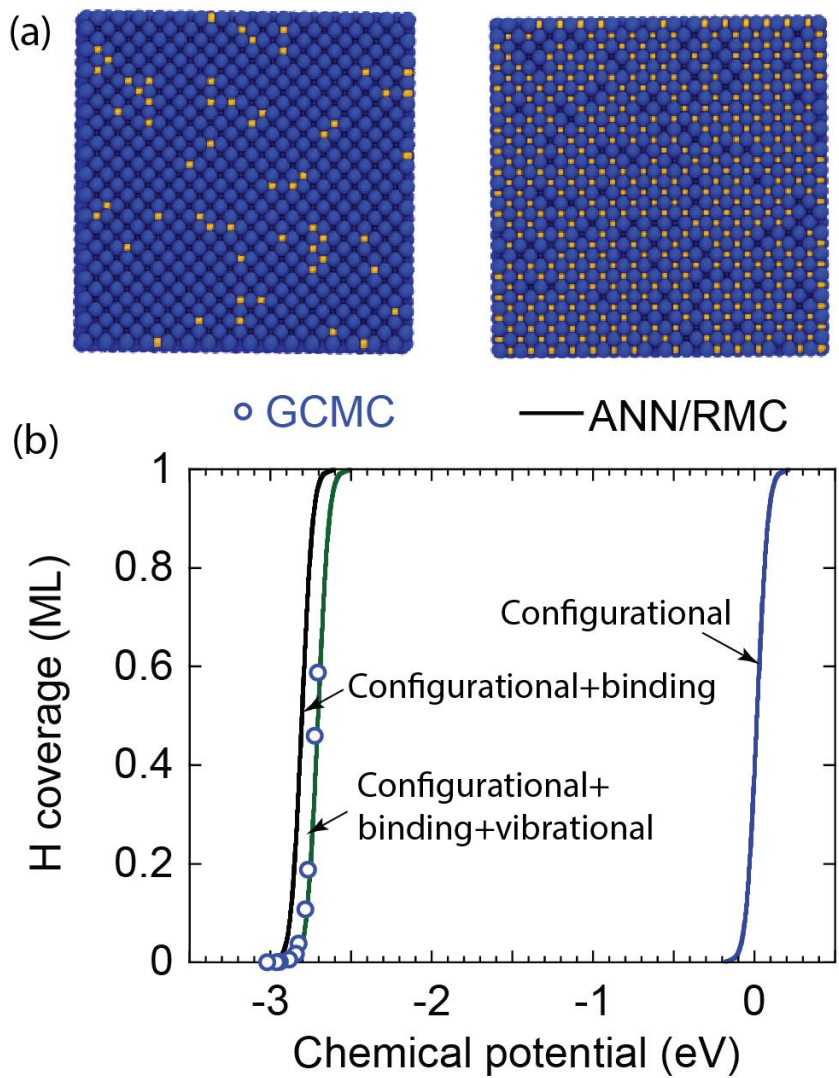

Figure 13: (a) Adsorbed H on Ni (100) surface at 298 K. Blue: Ni; Orange: H. (b) Adsorption isotherm from GCMC (circle) and ANN/RMC (solid lines). For ANN/RMC results with different terms are shown.

It can be shown that for the off-lattice system,

$$\Delta\mu = \mu_{binding} + \mu_{vibrational} + \Delta\mu_{configurational}. \tag{14}$$

$\Delta\mu_{configurational}$ is calculated using Equation (2) or (4). $\mu_{binding}$ arises from H binding to the Ni surface, while $\Delta\mu_{vibrational}$ arises due to the vibrational motion of the H atoms. These terms are calculated as

$$\mu_{binding} = V_i \tag{15}$$

and

$$F_{vib} = \sum_{i=1}^{3N}\left[\frac{h\omega_i}{2} + kT\ln(1 - e^{-h\omega_i/kT})\right] \tag{16}$$

$$\mu_{vibrational} = \frac{dF_{vib}}{dn}$$

In Equation (16), $\omega_i$ is the vibrational frequency obtained using normal mode analysis and $h$ is the Planck's constant. The vibrational frequencies were determined using from the dynamical matrix. $F_{vib}$ was found to be linear with respect to the number of adsorbed H atoms $n$.

Figure 13 shows the adsorption isotherm obtained by including the different terms. ANN/RMC results are shown in lines, whereas GCMC result is shown in circles. Section S6 of Supporting Information provides the same data in table form. When only the configurational part is included the isotherm is shifted to the right. Including the binding energy term brings the isotherm closer to the GCMC result. However, the inclusion of vibrational term ($\mu_{vibrational} = 0.105\ eV$) is essential to match with GCMC. The adsorption behaviour is correctly obtained with the ANN/RMC framework when all terms are included.

## 5. Conclusions

This paper demonstrates application of ANN models in RMC-based thermodynamic calculations. Previously, the RMC-based method required a database/library of configurations for interpolation, so that probability of local atomic arrangements can be obtained. The use of ANN model brings several advantages. First, the database no longer needs to be retained after the model has been fitted, which is the main motivation for this work. The compactness of the ANN model makes implementation of the thermodynamic calculations easier. The MATLAB codes provided as Supporting Information demonstrate these aspects. The ANN model solves a major issue in statistical thermodynamics that analytical, differentiable, closed-form expressions for the probability distributions are often not available. Moreover, as shown in Figure 4 the ANN model is more accurate than the interpolation method while requiring less data. ANN/RMC approach is found to be efficient and accurate in predicting the chemical potentials and adsorption isotherm. We have also extended our approach to off-lattice systems, namely, the H adsorption isotherm on 4-fold hollow sites of Ni (100) surface.

The ANN based approach presented here lays the foundation for extension to more complex lattice structures and longer-ranged interactions, which will be explored in future. There are situations where the single SRO parameter studied here may not be adequate for the problem in question. For example, repulsive lateral interactions can cause the adsorbed species to form ordered structures. It was shown in Ref. [29] that three SRO parameters are required for ordered structures in the NiPt system. Similarly, multiple order parameters are required in case of multiple site types or multisite occupation[64–68].

One limitation of the present approach is regarding the choice of sample points for building the training set, viz., which part of the $x - z$ space to sample and how many points. This is linked to the uncertainty in the predictions. Unlike other ML techniques such as Gaussian process regression (GPR)[52], uncertainty estimates are not available with the present approach. In future, attempts will be made to include the uncertainty using GPR, kriging or Kalman filter techniques. Samples can be added where the uncertainty is high, which could benefit the ANN model construction. Direct use of GPR for thermodynamic calculations may not be advisable as GPR will require dataset at all times, which is contrary to the original goal of this work. Moreover, in GPR the computational time scales as $s^3$, where $s$ is the number of samples. The high runtime in GPR is attributed to the inverse calculation of large covariance matrix. In the present study, $s \approx 400$. This number can be easily higher when we consider more complex lattice structures with multiple SRO parameters. Thus, it is hoped that neural networks might continue to be an efficient and reliable tool for training probability distribution functions with more complex problems.

## 6. Data and Software Availability

Software (MATLAB source code) is freely available at:

https://sites.google.com/view/abhijitchatterjee/download.

More description in provided in the Supporting Information.

## 7. Author contributions

AKB performed RMC calculations, training of the ANN models, GCMC calculations for the lattice systems and was involved in the creation of figures and writing of the manuscript. SR performed

GCMC calculations for the off-lattice system and normal mode analysis. GA was involved in the interpolation method. AC was involved in conceptualization, supervision, development of methodology, writing, reviewing and editing.

## 8. Acknowledgements

AC acknowledges support from Science and Engineering Research Board, Grant Nos. EMR/2017/001520 and MTR/2019/000909, and National Supercomputing Mission DST/NSM/R&D_HPC_Applications/2021/02.

**TOC image**

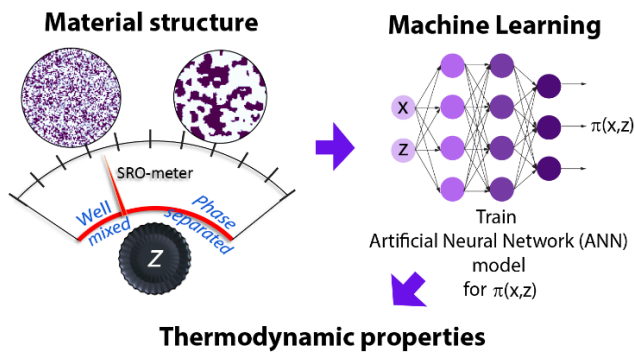